\documentclass[conference]{IEEEtran}
\usepackage{amssymb,amsmath,amsfonts}
\usepackage{color}
\usepackage{mathtools}
\usepackage{mathrsfs}
\usepackage{bm}
\usepackage{graphicx}		
\usepackage{epstopdf}		
\usepackage[font={small}]{caption}
\usepackage{caption}
\usepackage{subcaption}
\usepackage{multicol}
\usepackage{enumerate}
\usepackage{units}
\usepackage[numbers]{natbib}
\usepackage{algorithm}
\usepackage{flushend}

\DeclareMathOperator*{\argmax}{arg\,max}

\DeclarePairedDelimiter\abs{\lvert}{\rvert}

\begin{document}

\title{On the Ergodic Rate of OFDMA Cognitive Radios \\ under Imperfect Cross-Link Information}

\author{\IEEEauthorblockN{H. Saki, A. Shojaeifard, M. M. Mahyari, and M. Shikh-Bahaei}
\IEEEauthorblockA{Institute of Telecommunications, King's College London, London, WC2R 2LS, United Kingdom\\
Email: \{hadi.saki,a.shojaeifard,m.sbahaei\}@kcl.ac.uk}}

\maketitle

\begin{abstract}

The ergodic rate performance and limits of orthogonal frequency-division multiple access (OFDMA) cognitive radios (CRs) is studied under imperfect cross-link knowledge. We propose a novel stochastic interference management and exploitation technique to mitigate and control the imposed interference by CRs on licensed users in underlay spectrum sharing. The optimum downlink channel-adaptive resource allocation (RA) algorithm is designed to maximize the CRs functionality subject to satisfying average transmit and probabilistic peak interference power constraints. An expression for the cumulative density function (cdf) of CRs' received signal-to-interference-plus-noise ratio (SINR) is developed to evaluate the resultant ergodic rate. Simulation studies are conducted to examine the proposed RA algorithm and investigate the impact of parameters uncertainty on the overall system performance. 

\end{abstract}

\section{Introduction}
Spectrum sharing has attracted a lot of attention recently as measurements performed by regulators have indicated that the allocated radio spectrum 
experiences low utilization \cite{4812028}. In fact, according to the Office of Communications (Ofcom) report \cite{7678715}, the spectrum is severely underutilized in both spacial and temporal dimensions as a result of static management policies, and that 90\% of locations have around 100 MHz of spectrum available for other services. Cognitive radio (CR), defined as an intelligent system capable of observing the surrounding environment and adapting accordingly \cite{788210}, is a promising solution to overcome the spectrum crunch in modern wireless systems by exploiting spectral opportunities.

Three main paradigms have been proposed in regards to CRs access to primary spectrum \cite{4481339}, \cite{6118255}: (i) underlay approach where CRs coexist with primary users (PUs), 
provided they satisfy an interference limit set by a regulatory authority (ii) overlay approach in which CRs access the vacant parts of the licensed spectrum, and (iii) hybrid approach, a combination of the two former strategies where CRs can dynamically select between the underlay and overlay modes depending on the traffic and interference characteristics. Here, we focus on underlay spectrum access, where interference management and exploitation is essential for achieving desirable performance whilst tackling any harmful cross-service interference. 

Orthogonal frequency-division multiple-access (OFDMA) has emerged as a prominent air interface technology for new generation of wireless communication systems including long term evolution (LTE)-advanced \cite{995857,neh} and \cite{4027580}. Further, OFDMA is considered a de facto standard for CR networks due to its inherent advantages in terms of flexibility and adaptability in allocating radio resources in shared-spectrum environments \cite{5510778}. Resource allocation (RA) plays a significant role in improving the spectral efficiency of conventional OFDMA systems \cite{kob,SakiS15,4382916}. In addition, RA is an active area of research in OFDMA CR networks with the aim of achieving a balance between maximizing the cognitive system performance and minimizing the inflicted interference on the licensed users.  

Analysing the performance and limits of CR is essential towards identifying the viable business models and initiating standardization. A substantial amount of study on various performance metrics of CR systems has recently emerged, see, e.g., \cite{5336868}, \cite{4786456}, and the references herein. A central assumption for controlling the cross-service interference in many of the works in the literature is perfect knowledge of the interfering links between the CRs and licensed receivers. Imposing transmit and interference power constraints under the perfect channel state information (CSI) assumption guarantees a limited inflicted interference on the PUs. However, obtaining cross-link information requires a cooperative signalling protocol between the primary and cognitive networks \cite{5967979}. This may be infeasible in practice, and even if a cross-connection mechanism is implemented, the inherent nature of wireless environment and estimation errors result in noisy CSI. 

A number of works on CR have considered the impact of imperfect cross-link knowledge: \cite{5336868} studies the rate under an average interference constraint, the capacity under peak interference constraint is derived in \cite{5419086}, and optimum power allocation policy and ergodic capacity are derived under two different interference outage and signal-to-interference-plus-noise ratio (SINR) outage constraints in \cite{6185693}. However, all of these studies consider a single CR scenario, and further, \cite{5336868} and \cite{5419086} adopt deterministic interference constraints. 

The motivation of this work is to analyze the impact of noisy cross-link knowledge on ergodic transmission rate of multi-user multi-band OFDMA CR networks under average transmit and probabilistic peak interference power constraints. We propose a novel stochastic approach for mitigating the CSI imperfections and design the optimum downlink RA algorithm in the cognitive network. Further, in order to compute the ergodic rate, we develop an approximated expression for the cumulative density function (cdf) of CRs' received SINR.                  

\textit{Notations:} $\mathscr{E}\{x\}$ and $\mathscr{P}(x)$ denote the expectation and probability of $x$, respectively. $var(x)$ denotes the variance of $x$ and $cov(y,z)$ is defined as the covariance of $y$ and $z$. 
$[x]^{+}$ signifies $\max(0,x)$, and $x^{*}$ is the optimal value of $x$.

\section{Cognitive System Model and Preliminaries}

\begin{figure}[t]
\centering
\includegraphics[width=.325\textwidth]{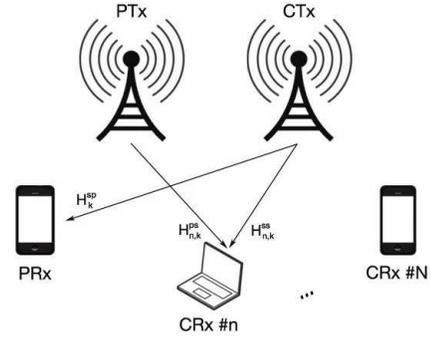}
\caption{Schematic diagram of the underlay spectrum sharing system. For simplicity, the corresponding channels of a single CR are drawn.}
\vspace*{-1em}
\label{CRmodel}
\end{figure}

Consider a shared-spectrum environment as shown in Fig. \ref{CRmodel}, where a secondary network of $N$ CRs (indexed by $n$) coexist with a primary network over $K$ non-overlapping bands (indexed by $k$) subject to satisfying the power constraints set by a regulatory authority. The bandwidth of each band is assumed to be much smaller than the coherence bandwidth of the wireless channel, thus, each subcarrier experiences frequency-flat fading. Let $H^{ss}_{n,k}(t)$, $H^{ps}_{n,k}(t)$, and $H^{sp}_{k}(t)$, at time $t$, denote the complex channel gains over band $k$ from the cognitive transmitter (CTx) to $n^{\text{th}}$ cognitive receiver (CRx), primary transmitter (PTx) to $n^{\text{th}}$ CRx, and CTx to primary receiver (PRx), respectively. The channel power gains $|H^{ss}_{n,k}(t)|^2$, $|H^{ps}_{n,k}(t)|^2$, and $|H^{sp}_{k}(t)|^2$, are assumed to be ergodic and stationary with continuous probability density functions (pdfs) $f_{|H^{ss}_{n,k}|^2}(.)$, $f_{|H^{ps}_{n,k}|^2}(.)$, and $f_{|H^{sp}_{k}|^2}(.)$, respectively. In addition, the instantaneous values and distribution information of secondary-secondary channel gains are assumed to be available at the CTx \cite{6185693}. Due to the impact of several systematic factors, such as channel estimation error, feedback delay, and mobility, perfect cross-link information is not available. We consider imperfect cross-link knowledge at the CTx obtained from maximum likelihood (ML) estimation and model the associated uncertainty in cross-link in the following form
\begin{align}
H^{sp}_{k}(t) = \hat{H}^{sp}_{k}(t) + \Delta H^{sp}_{k}(t)
\label{errorF}
\end{align}
where over subcarrier (band) $k$, at time $t$, $H^{sp}_{k}(t)$ is the true cross-link gain, $\hat{H}^{sp}_{k}(t)$ is the estimated channel gain considered to be known, and $\Delta H^{sp}_{k}(t)$ denotes the estimation error. $H^{sp}_{k}(t)$, $\hat{H}^{sp}_{k}(t)$, and $\Delta H^{sp}_{k}(t)$ are assumed to be zero-mean complex Gaussian random variables with respective variances $\delta^{2}_{H^{sp}_{k}}$, $\delta^{2}_{\hat{H}^{sp}_{k}}$, and $\delta^{2}_{\Delta H^{sp}_{k}}$ \cite{5419086}. For robust receiver design, we consider the estimation $\hat{H}^{sp}_{k}(t)$ and error $\Delta H^{sp}_{k}(t)$ to be statistically correlated random variables with a correlation factor $\rho$, which determines the accuracy of channel estimation in relation to true channel states, defined as \cite{4712722}
\begin{align}
\rho = \sqrt{\delta^{2}_{\Delta H^{sp}_{k}} \Big/ \left( \delta^{2}_{\Delta H^{sp}_{k}} + \delta^{2}_{H^{sp}_{k}} \right)}. 
\end{align} 
We proceed by deriving posteriori distributions of the true cross-link given the estimation and hence the estimation error given the estimation. For notational brevity, we omit the time reference $t$ where the context is clear. 

\textit{Proposition 1:}
The posterior distribution of the estimation error $\Delta H^{sp}_{k}$ given the estimation $\hat{H}^{sp}_{k}$ is a complex Gaussian random variable with respective mean and variance of
\begin{gather}
\mu_{\Delta H^{sp}_{k} | \hat{H}^{sp}_{k}} = \mathscr{E}(\Delta H^{sp}_{k} | \hat{H}^{sp}_{k}) = \mathscr{E}(\Delta H^{sp}_{k}) + \nonumber \\ \frac{cov(\Delta H^{sp}_{k},\hat{H}^{sp}_{k})}{\delta^{2}_{\Delta H^{sp}_{k}} + \delta^{2}_{H^{sp}_{k}}} \left( \hat{H}^{sp}_{k} - \mathscr{E}(\hat{H}^{sp}_{k}) \right) = \rho^{2} \hat{H}^{sp}_{k} 
\label{ExpDaltaHhat}
\end{gather}
\begin{gather}
\delta^2_{\Delta H^{sp}_{k} | \hat{H}^{sp}_{k}} = var(\Delta H^{sp}_{k} | \hat{H}^{sp}_{k}) \nonumber \\ = \delta^{2}_{\Delta H^{sp}_{k}} \biggr[1 - \frac{cov^{2}(\Delta H^{sp}_{k},\hat{H}^{sp}_{k})}{\delta^{2}_{\Delta H^{sp}_{k}} \, \delta^{2}_{\hat{H}^{sp}_{k}}} \biggr] = (1 - \rho^{2}) \delta^{2}_{\Delta H^{sp}_{k}}.
\end{gather}

\textit{Proposition 2:} 
The posterior distribution of the true cross-link $H^{sp}_{k}$ given the estimation $\hat{H}^{sp}_{k}$ is a complex Gaussian random variable with respective mean and variance of
\begin{gather}
\mu_{H^{sp}_{k} | \hat{H}^{sp}_{k}} = \mathscr{E}(\hat{H}^{sp}_{k} + \Delta H^{sp}_{k} | \hat{H}^{sp}_{k}) \nonumber \\ = \mathscr{E}(\hat{H}^{sp}_{k} | \hat{H}^{sp}_{k})  + \mathscr{E}(\Delta H^{sp}_{k} | \hat{H}^{sp}_{k}) = (1 + \rho^2) \hat{H}^{sp}_{k}  
\end{gather}
\begin{gather}
\delta^2_{H^{sp}_{k} | \hat{H}^{sp}_{k}} = var(\hat{H}^{sp}_{k} + \Delta H^{sp}_{k} | \hat{H}^{sp}_{k}) = var(\hat{H}^{sp}_{k} | \hat{H}^{sp}_{k}) \nonumber \\ + var(\Delta H^{sp}_{k} | \hat{H}^{sp}_{k}) + 2 cov(\Delta H^{sp}_{k} | \hat{H}^{sp}_{k},\hat{H}^{sp}_{k} | \hat{H}^{sp}_{k}) \nonumber \\ = (1 - \rho^2) \delta^{2}_{\Delta H^{sp}_{k}}.
\end{gather}

In the OFDMA CR network, the downlink RA algorithm allocates power and subcarrier to CRs based on the fading-induced fluctuations in the true secondary-secondary and estimated interfering secondary-primary channels. Each subcarrier is assigned exclusively to at most one CR at any given time, hence, there is no mutual interference between different CRxs \cite{4027580}. It should also be noted that by utilizing an appropriate cyclic prefix (CP), the inter-symbol-interference (ICI) can be ignored \cite{5967979}. The instantaneous received SINR of CR $n$ over band $k$ with a fixed transmit power $P_{n,k}$ is given by
\begin{align}
\gamma_{n,k} = \frac{P_{n,k} |H^{ss}_{n,k}|^2}{ \sigma^2_{n} + \sigma^2_{ps} }
\end{align}
where $\sigma^{2}_{n}$ is the power of circularly symmetric zero-mean complex Gaussian noise and $\sigma^{2}_{ps}$ is the received interference power from the primary network taken as white Gaussian noise \cite{5967979}, \cite{5290301},
Without loss of generality, $\sigma^{2}_{n}$ and $\sigma^{2}_{ps}$ are assumed to be the same across all users and subcarriers \cite{5290301}, \cite{5672619}. 

\section{Interference Management and Exploitation}

In a spectrum sharing paradigm, and particularly for delay-sensitive licensed services, the PUs' quality of service (QoS) is highly dependent on the instantaneous received interference power from the CRs. To protect the licensed spectrum from harmful interference, extra constraints are needed to facilitate interference control. However, due to the underlying uncertainties about the shared-spectrum environment and PUs' activity, it is unrealistic to assume that the CRs can always satisfy deterministic interference constraints. In practice, the tolerable interference level is confined by a maximum collision probability that guarantees a certain grade of QoS for PUs. 

In this paper, we consider an underlay spectrum sharing system where the primary network tolerates a maximum collision probability of $\varepsilon$. Collision is considered to occur when the interference level imposed by CRs is higher than $I^{th}$. Adopting a probabilistic interference constraint is crucial for robust interference management given noisy cross-link knowledge. To improve overall system performance and to mitigate the
impact of channel estimation errors, the following probabilistic peak interference constraint is considered
\begin{gather}
\mathscr{P} \left( \sum_{n=1}^{N} \sum_{k=1}^{K} \varphi_{n,k}(\gamma_{n,k}) P_{n,k}(\gamma_{n,k}) \big| H^{sp}_{k} \big| \hat{H}^{sp}_{k} \big|^2 \! > \! I^{th} \right) \! \leq \! \varepsilon 
\label{IMeq2}
\end{gather}
where for cognitive user $n$ over band $k$, $\varphi_{n,k}(\gamma_{n,k})$ is the time-sharing factor (subcarrier allocation policy), and $P_{n,k}(\gamma_{n,k})$ is the allocated transmit power. Assuming equal variance $\delta^2_{H^{sp}_{k} | \hat{H}^{sp}_{k}}$ across all users and subcarriers, the collision probability constraint in (\ref{IMeq2}) can be expressed as
\begin{gather}
\mathscr{P} \left(\! \delta^2_{H^{sp}_{k} | \hat{H}^{sp}_{k}} \sum_{n=1}^{N}  \sum_{k=1}^{K} \varphi_{n,k}(\gamma_{n,k}) P_{n,k}(\gamma_{n,k}) |\Xi_{k}|^2 \! > \! I^{th} \! \right) \! \leq \! \varepsilon
\label{CHAINOTNOT}
\end{gather}
where $|\Xi_{k}|^2$ is a non-central Chi-Square random variable with two degrees of freedom and non-centrality parameter $\mu_{\Xi_{k}} = \abs[\Big]{ \mu_{H^{sp}_{k} | \hat{H}^{sp}_{k}} / \delta_{H^{sp}_{k} | \hat{H}^{sp}_{k}} } ^2$. Note that (\ref{CHAINOTNOT}) includes a sum of non-equal-weighted Chi-Square random variables. 
In general, obtaining the exact distribution of the linear combination of weighted Chi-Square random variables is rather complex. Although several approximations have been proposed, e.g., \cite{1987}, they mostly rely on prior knowledge of the weights, in this case prior knowledge of power and subcarrier assignments. Hence, conventional approaches cannot be applied to the resource allocation problem under consideration. In this work, we propose a novel simple approximation based on the moments of $\delta^2_{H^{sp}_{k} | \hat{H}^{sp}_{k}} \sum_{n=1}^{N} \! \sum_{k=1}^{K} \! \varphi_{n,k}(\gamma_{n,k}) P_{n,k}(\gamma_{n,k}) |\Xi_{k}|^2$. 
Let $\beta_{k} = \sum_{n=1}^{N} \delta^2_{H^{sp}_{k} | \hat{H}^{sp}_{k}} \varphi_{n,k}(\gamma_{n,k}) P_{n,k}(\gamma_{n,k})$, hence
\begin{align}
\delta^2_{H^{sp}_{k} | \hat{H}^{sp}_{k}} \sum_{n=1}^{N} \sum_{k=1}^{K} \! \varphi_{n,k}(\gamma_{n,k}) P_{n,k}(\gamma_{n,k}) |\Xi_{k}|^2 \! = \! \sum_{k=1}^{K} \! \beta_{k} |\Xi_{k}|^2.
\label{ProbInt}
\end{align} 

\begin{figure*}[ht]
\begin{minipage}[b]{0.49\linewidth}
\centering
\includegraphics[width=\textwidth]{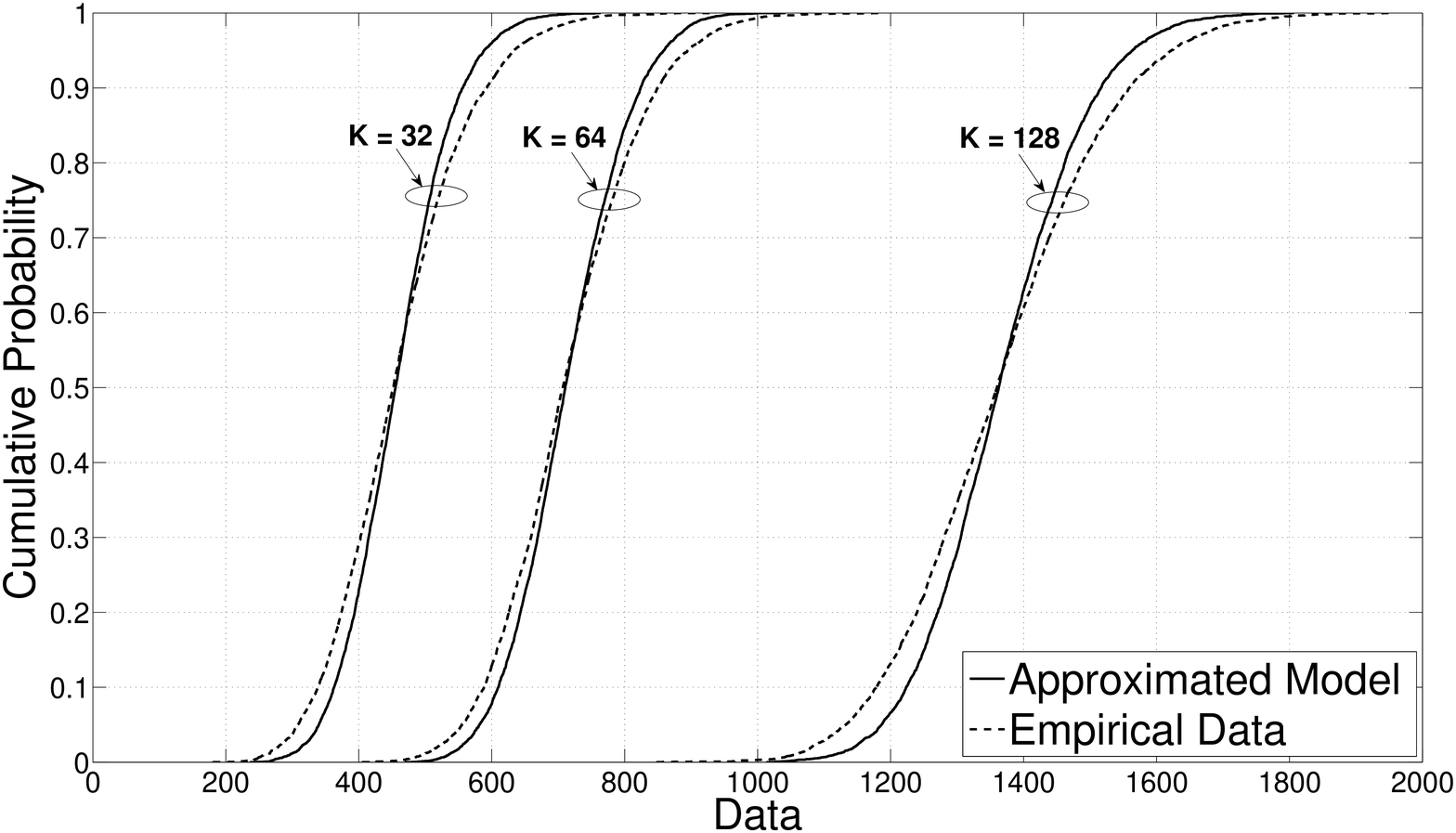}
\subcaption{$\beta_{k} \sim \text{Chi-Square}(2,2)$, $\delta^{2}_{H^{sp}_{k} | \hat{H}^{sp}_{k}}=1$.}
\label{fig:figure1}
\end{minipage}
\hfill
\begin{minipage}[b]{0.49\linewidth}
\centering
\includegraphics[width=\textwidth]{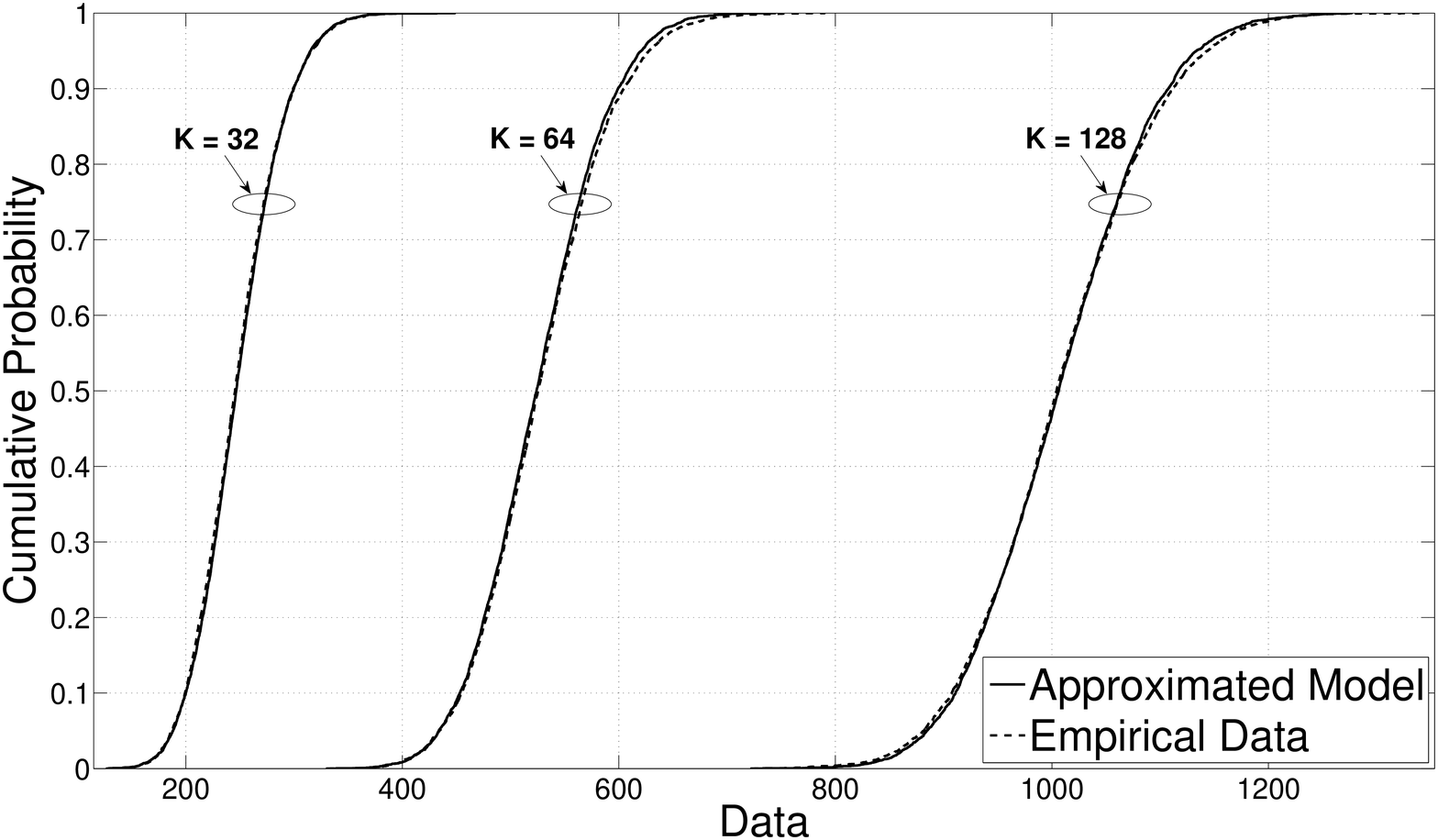}
\subcaption{$\beta_{k} \sim \text{Gamma}(2,0.5,4)$, $\delta^{2}_{H^{sp}_{k} | \hat{H}^{sp}_{k}}=0.5$.}
\label{fig:figure2}
\end{minipage}
\centering{}\caption{Approximated Model and Empirical Data cdfs, obtained from Monte-Carlo simulations.
\label{cdf-1-1}}
\end{figure*}

\textit{Proposition 3:} The distribution of the sum of non-equal-weighted non-central Chi-Square random variables, $\sum_{k=1}^{K} \beta_{k}|\Xi_{k}|^2$, is similar to that of a weighted non-central Chi-Square-distributed random variable $\xi \chi_{D}^{2}(\delta^{'})$, where $\delta^{'}$, $D$, and $\xi$ are respectively the non-centrality parameter, degrees of freedom, and weight of the new random variable:
\begin{align}
\delta^{'} = \sum_{k=1}^{K} \mu_{\Xi_{k}} \label{APsub1}\\
D = 2K \label{APsub2}\\
\xi = \frac{\sum_{k=1}^{K} \beta_{k} (2 + \mu_{\Xi_{k}})}{2K + \sum_{k=1}^{K} \mu_{\Xi_{k}}}. \label{APsub3}
\end{align}
To investigate the above similarity, we compare the cdf of the proposed Chi-Square distribution with that of (\ref{ProbInt}), using Monte-Carlo simulations. The results in Fig. \ref{cdf-1-1} illustrate that the approximation is accurate over a wide range of practical values for $K$ over randomly-distributed - e.g., Chi-Square or Gamma - weights $\beta_{k}$. Now (\ref{CHAINOTNOT}) can be simplified to:
\begin{align}
& \mathscr{P} \Biggr( \delta^2_{H^{sp}_{k} | \hat{H}^{sp}_{k}} \sum_{n=1}^{N} \sum_{k=1}^{K} \varphi_{n,k}(\gamma_{n,k}) P_{n,k}(\gamma_{n,k}) |\Xi_{k}|^2 > I^{th} \Biggr) \nonumber \\ & \approx \mathscr{P} \left( \xi \chi_{D}^{2}(\delta^{'}) > I^{th} \right).
\end{align}
Given the non-centrality parameter is small relative to the degrees of freedom, we can approximate the non-central Chi-Square distribution with a central one using the following \cite{1987}
\begin{align}
\mathscr{P} \left( \xi \chi_{D}^{2}(\delta^{'}) > I^{th} \right) \approx \mathscr{P} \left( \chi_{D}^{2}(0) > \frac{I^{th}/\xi}{1 + \delta^{'}/D} \right).
\label{PrNCPrC}
\end{align}
The right hand side of (\ref{PrNCPrC}) can be formulated using the upper Gamma function as 
\begin{align}
\mathscr{P} \left( \chi_{D}^{2}(0) > \frac{I^{th}/\xi}{1 + \delta^{'}/D} \right) = \frac{G(K,\frac{I^{th}/\xi}{2(1 + \delta^{'}/D)})}{G(K)}
\end{align}
where $G(.,.)$ is the upper incomplete Gamma function, and $G(.)$ is the complete Gamma function.

\textit{Proposition 4:} For all integer values $K \neq 1$, and all positive $\frac{I^{th}/\xi}{1 + \delta^{'}/D}$ - this condition is always true because, $I^{th}$, $\delta^{'}$ $\beta_{k}$, and $K$ are positive, consequently, $\xi$, $\delta^{'}$, and $D$ are also positive - the deterministic inequality 
\begin{gather}
\delta^2_{H^{sp}_{k} | \hat{H}^{sp}_{k}} \sum_{k=1}^{K} (2 + \mu_{\Xi_{k}}) \sum_{n=1}^{N} \varphi_{n,k}(\gamma_{n,k}) P_{n,k}(\gamma_{n,k}) \nonumber \\ \leq \frac{K \, I^{th}}{(K!)^{1/K} \ln \left( 1 - (1 - \varepsilon)^{1/K} \right)} 
\label{DetermIC}
\end{gather}
satisfies the probabilistic inequality (\ref{CHAINOTNOT}). Therefore, the constraint (\ref{CHAINOTNOT}) can be replaced by (\ref{DetermIC}).

Additionally, mitigating the interference between neighbouring cells is a vital issue due to the increasing frequency reuse aggressiveness in modern wireless communication systems \cite{5770666}. As a remedy to inter-cell interference, and to maintain effective and efficient power consumption, we impose a total average transmit power limit $P_{t}$ on the cognitive network:
\begin{align}
\sum_{n=1}^{N} \sum_{k=1}^{K} \mathscr{E}_{\gamma_{n,k}} \Big\{ \varphi_{n,k}(\gamma_{n,k}) P_{n,k}(\gamma_{n,k}) \Big\} \leq P_t.
\label{AvePowerConst}
\end{align}

\section{Ergodic Rate}

The ergodic transmission rate (bps/Hz) of the OFDMA CR network operating under interference and transmit power constraints can be expressed as
\begin{align}
R = \sum_{n=1}^{N} \sum_{k=1}^{K} \mathscr{E}_{\gamma_{n,k}} \left\{ \log_{2} \left(1 \! + \! \frac{\gamma_{n,k} P_{n,k}(\gamma_{n,k}) }{\min \left(\frac{P_t}{K} , \frac{I_{t}}{N^{sp}} \right)} \varphi_{n,k}(\gamma_{n,k}) \right) \right\}
\label{AASE}
\end{align} 
where $N^{sp} = \delta^2_{H^{sp}_{k} | \hat{H}^{sp}_{k}} \sum_{k=1}^{K} (2 + \mu_{\Xi_{k}})$ and $I_{t} = K \, I^{th} \big/ (K!)^{1/K} \ln \left( 1 - (1 - \varepsilon)^{1/K} \right)$. In order to evaluate the ergodic rate, the distribution of CRs' received SINR, dependent on secondary-secondary and secondary-primary channel gains, must be developed. Limited by the constraints (\ref{DetermIC}) and (\ref{AvePowerConst}), the cdf of $\gamma_{n,k}$ can be written as
\begin{align}
F_{\gamma_{n,k}}(\Gamma) = \mathscr{P} \biggr( \frac{P_{t} |H^{ss}_{n,k}|^2}{K (\sigma^2_{n} + \sigma^2_{ps})} \leq \Gamma , \frac{I_{t} |H^{ss}_{n,k}|^2}{N^{sp} (\sigma^2_{n} + \sigma^2_{ps})} \leq \Gamma \biggr).
\label{cdfofG}
\end{align}
The expression in (\ref{cdfofG}) can be simplified by considering the cases $\frac{P_{t} |H^{ss}_{n,k}|^2}{K (\sigma^2_{n} + \sigma^2_{ps})} \lesseqqgtr \frac{I_{t} |H^{ss}_{n,k}|^2}{N^{sp} (\sigma^2_{n} + \sigma^2_{ps})}$ and conditioning on $N^{sp}$:
\begin{align}
& 1 - \mathscr{P} \biggr( \frac{P_{t} |H^{ss}_{n,k}|^2}{K (\sigma^2_{n} + \sigma^2_{ps})} > \Gamma , \frac{I_{t} |H^{ss}_{n,k}|^2}{N^{sp} (\sigma^2_{n} + \sigma^2_{ps})} > \Gamma \biggr) = \nonumber \\ & 1 -
  \begin{dcases}
   \mathscr{P} \biggr( |H^{ss}_{n,k}|^2 > \frac{K \Gamma (\sigma^2_{n} + \sigma^2_{ps})}{P_{t}} \biggr) & N^{sp} \leq \frac{I_{t} K}{P_t} \nonumber \\
   \mathscr{P} \biggr( |H^{ss}_{n,k}|^2 > \frac{N^{sp} \Gamma (\sigma^2_{n} + \sigma^2_{ps})}{I_{t}} \biggr) & N^{sp} > \frac{I_{t} K}{P_t}.
  \end{dcases}\\
  \label{sigmaOPT}
\end{align}
Invoking central limit theorem for large values of K, $N^{sp}$ can be approximated as a Gaussian random variable with mean $\mu_{N^{sp}} = 2K \delta^2_{\hat{H}^{sp}_{k}} \left( 1 + \rho^2 \right)^2 + 2 K \left( 1 - \rho^2 \right)^2 \delta^2_{\Delta H^{sp}_{k}}$ and variance $\delta^{2}_{N^{sp}} = 4K \delta^4_{\hat{H}^{sp}_{k}} \left( 1 + \rho^2 \right)^4$. 
Suppose that $|H^{ss}_{n,k}|^2$ follows an Exponential distribution with mean $\mu_{|H^{ss}_{n,k}|^2}$, hence, with further manipulation,
a closed-form expression for cdf of $\gamma_{n,k}$ is developed in (\ref{CDFofgamma}). Trivially, through respective differentiation of (\ref{CDFofgamma}), the pdf of $\gamma_{n,k}$ can be obtained. 

\section{Radio Resource Allocation Algorithm}

\begin{figure*}[!t]
\normalsize
\setcounter{equation}{21}
\begin{align}
& F_{\gamma_{n,k}}(\Gamma) \approx \nonumber \\ & 1 - \frac{1}{2} \exp \left(\! \frac{- K \Gamma (\sigma^2_{n} + \sigma^2_{ps})}{P_t \mu_{|H^{ss}_{n,k}|^2}} \! \right) \Biggr[1 + erf \biggr( \frac{\frac{I_{t} K}{P_t} - \mu_{N^{sp}}}{\sqrt{2 \delta^{2}_{N^{sp}}}} \biggr) \Biggr] - \frac{1}{2} \exp \Biggr(\! {\frac{\Gamma (\sigma^2_{n} + \sigma^2_{ps})  \left( -2 \mu_{N^{sp}} \mu_{|H^{ss}_{n,k}|^2} I_{t} + \delta^{2}_{N^{sp}} \Gamma (\sigma^2_{n} + \sigma^2_{ps}) \right)}{2 \mu^2_{H^{ss}_{n,k}} {I_{t}}^2}} \! \Biggr) \nonumber \\ & \Biggr[ 1 - erf \biggr( \frac{ \mu_{|H^{ss}_{n,k}|^2} I_{t} \left( - \mu_{N^{sp}} + \frac{I_{t} K}{P_{t}} \right) + \delta^{2}_{N^{sp}} \Gamma (\sigma^2_{n} + \sigma^2_{ps})}{\sqrt{2} \mu_{|H^{ss}_{n,k}|^2} I_{t} \delta_{N^{sp}}} \biggr) \Biggr].
\label{CDFofgamma}    
\end{align}
\hrulefill
\vspace*{1pt}
\end{figure*}

In this section, we design the optimal downlink power and subcarrier allocation algorithm to maximize the ergodic rate of the OFDMA CR network subject to satisfying the transmit and interference power constraints under noisy cross-link CSI. \linebreak The optimization problem can be formulated as  
\begin{subequations}
\begin{gather}
\max_{\varphi_{n,k}(\gamma_{n,k}),P_{n,k}(\gamma_{n,k})} \;\; R  \label{OF3a}\\ 
\text{s. t.:} \quad \text{constraints in (\ref{DetermIC}), (\ref{AvePowerConst})} \nonumber \\
\sum_{n=1}^{N} \varphi_{n,k}(\gamma_{n,k}) = 1 , \forall k \! \in \! \{1, ..., K\} \label{OF1d}\\
\varphi_{n,k}(\gamma_{n,k}) \in \{0,1\} , \forall n \in \{1, ..., N\} , \forall k \in \{1, ..., K\}  \label{OF1e}
\end{gather}
\end{subequations}
where constraints (\ref{OF1d}) and (\ref{OF1e}) ensure that every subcarrier is allocated to at most one CR. It can be observed that the optimization problem is convex with respect to $P_{n,k}(\gamma_{n,k})$, however, it is non-convex with respect to $\varphi_{n,k}(\gamma_{n,k})$. 
By applying Lagrangian dual decomposition, the non-convex optimization problem can be decomposed into independent sub-problems each corresponding to a given CR. 
By solving the Lagrangian optimization problem the following potential optimal power allocation solution can be obtained for user $n$ over subcarrier $k$ with Lagrangian multipliers $\mu$ and $\eta$
\begin{align}
& P^{*}_{n,k}(\gamma_{n,k}) = \nonumber \\ & \Biggr[\frac{1}{\ln(2) \! \left( \mu \! + \! \eta \delta^2_{H^{sp}_{k} | \hat{H}^{sp}_{k}} (2 + \mu_{\Xi_k}) \right)} \! - \! \frac{{\min \left(\frac{P_t}{K} , \frac{I_{t}}{N^{sp}} \right)}}{\gamma_{n,k}} \Biggr]^{+} \! . 
\label{Pstar}
\end{align}
The solution in (\ref{Pstar}) can be considered as a multi-level water-filling algorithm.
Using (\ref{Pstar}) and by applying the Karush-Kuhn-Tucker (KKT) conditions, the optimal subcarrier allocation problem is 
formulated as
\begin{align}
n^{*} = \argmax \Big( \Lambda(\gamma_{n,k}) \Big) \; , \; \forall n \in \{1,...,N\} \; , \; \forall k \in \{1,...,K\}\nonumber \\
\label{PROBLEMSCAPOPT}
\end{align}
where $n^{*}$ is the optimal CR index, and
\begin{align}
& \Lambda(\gamma_{n,k}) = \frac{\frac{\gamma_{n,k} P^{*}_{n,k}(\gamma_{n,k}) }{\min \left(\frac{P_t}{K} , \frac{I_{t}}{N^{sp}} \right)}}{\ln(2) \left( 1 + \frac{\gamma_{n,k} P^{*}_{n,k}(\gamma_{n,k})}{\min \left(\frac{P_t}{K} , \frac{I_{t}}{N^{sp}} \right)} \right)} + \nonumber \\ & \frac{\ln \left( 1 + \frac{\gamma_{n,k} P^{*}_{n,k}(\gamma_{n,k})}{\min \left(\frac{P_t}{K} , \frac{I_{t}}{N^{sp}} \right)} \right)}{\ln(2)}.
\label{SCAPOPT}
\end{align}
The optimal subcarrier allocation policy is achieved by assigning the $k^{th}$ subcarrier to the CR with the highest value of $\Lambda(\gamma_{n,k})$, i.e., placing a Lagrangian multiplier $\lambda_k$ between the first and second maximas of $\Lambda(\gamma_{n,k})$. If there are multiple equal maximas, the time-slot can be identically shared among the respective users. Here, we use the subgradient-based method to update the values of the multipliers $\mu$ and $\eta$:
\allowdisplaybreaks{
\begin{gather}
\mu^{i+1} = \mu^{i} - \tau^{i}_{1} \biggr( P_{t} - \sum_{n=1}^{N} \sum_{k=1}^{K} \varphi^{*}_{n,k}(\gamma_{n,k}) P^{*}_{n,k}(\gamma_{n,k}) \biggr) \label{SUB1} \\
\eta^{i+1}(\gamma_{n,k}) = \eta^{i}(\gamma_{n,k}) - \tau^{i}_{2} \; \times \nonumber \\  \left( I_{t} - \delta^2_{H^{sp}_{k} | \hat{H}^{sp}_{k}} \sum_{k=1}^{K} (2 \! + \! \mu_{\Xi_k}) \sum_{n=1}^{N} \varphi^{*}(\gamma_{n,k}) P^{*}_{n,k}(\gamma_{n,k}) \right) 
\label{SUBG}
\end{gather}
where for the iteration number $i$, $\tau^{i}_{1}$ and $\tau^{i}_{2}$ are the step sizes. The initial values of dual multipliers and step size selection are important towards obtaining the optimal solution, and can greatly affect the optimization problem convergence.}

\section{Performance Evaluation and Discussion}

The performance of the OFDMA CR network using the proposed RA algorithm, subject to satisfying average transmit and probabilistic peak interference power limits under imperfect cross-channel estimation, is studied. Perfect CSI is assumed between CTx and CRxs through an error-free feedback channel. $|H^{ss}_{n,k}|$, $\forall \{n,k\}$, are assumed to be Rayleigh-distributed and the secondary-secondary power gain mean values, $\mu_{|H^{ss}_{n,k}|^2}$, $\forall \{n,k\}$, are taken as Uniformly-distributed random variables within 0 to 2. 
Interfering cross-channel values, $H^{sp}_{k}$, $\forall \{k\}$, are distributed according to a complex Gaussian distribution with mean 0.05 and variance 0.1. The channel estimation and error for all sub-channels are taken as independent and identically distributed (i.i.d.) zero-mean Normally-distributed random variables. The noise and primary-secondary interference power spectral densities are set to $N_{0} = -110$ dBm and $2500 N_{0}$, respectively. All results correspond to a CR network with 64 subcarriers and 3 CRxs.  

\begin{figure}[t]
\includegraphics[width=.5\textwidth]{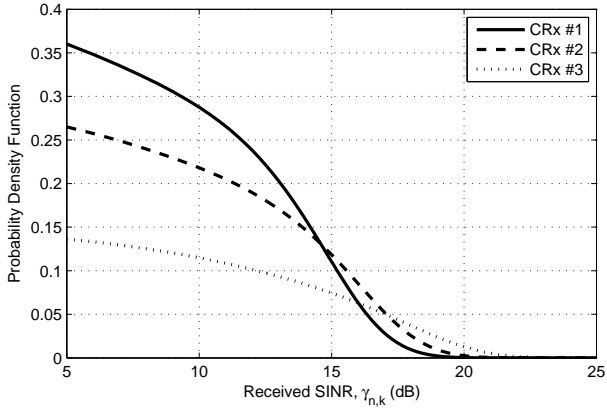} 
\caption{Probability density functions of the received SINR for OFDMA CRs in a given subcarrier $k$. System parameters are: $k = 55$, $P_t = 25$ Watts, $I^{th} = 5$ Watts, $\rho = 0.25$, $\varepsilon = 30\%$.}
\label{fig:PDFvsSINR}
\end{figure}

The approximated pdfs for the received SINR of the three CRs, dependent on noise floor and true secondary-secondary and estimated secondary-primary channel gains, in a randomly taken subcarrier, i.e., here $k=55$, is plotted in Fig. \ref{fig:PDFvsSINR}. The results are obtained using Monte-Carlo simulations. 

\begin{figure}
\includegraphics[width=.5\textwidth]{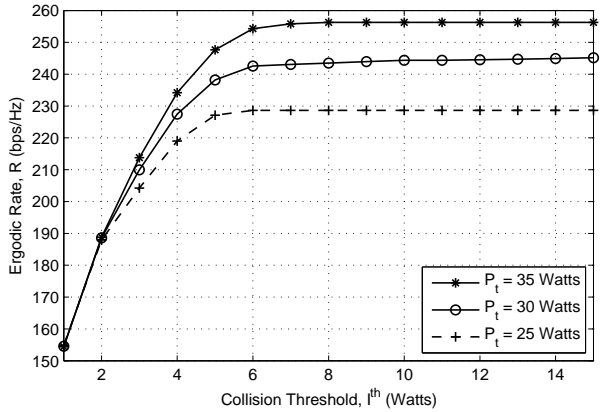} 
\caption{Ergodic rate versus the collision threshold with different values of average transmit power limit. System parameters are: $\rho = 0.5$, $\varepsilon = 15\%$.}
\label{fig:ERvsIth}
\end{figure}

Fig. \ref{fig:ERvsIth} shows the ergodic rate of the OFDMA CRs versus the collision threshold. As expected, a higher collision limit results in improved $R$ as $I^{th}$ limits the CRs' transmit power functionality. The gain in ergodic rate, however, approaches a plateau in high $I^{th}$ region as $P_t$ becomes the dominant power constraint. Given $P_t$ is dominant, imposing a higher average power setting enhances the CRs' uppermost performance, e.g., where $I^{th} = 10$ Watts, a 7.50\% increase in ergodic rate is realized as $P_t$ is increased from 20 to 25 Watts.  

\begin{figure}
\includegraphics[width=.5\textwidth]{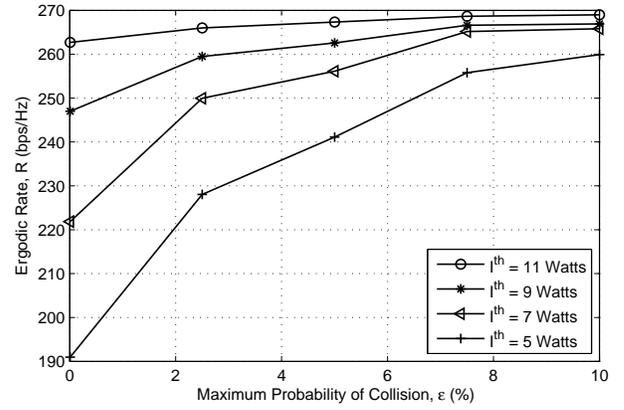} 
\caption{Ergodic rate against the probabilistic interference constraint with different values of collision threshold. System parameters are: $P_t = 40$ Watts, $\rho = 0.75$.}
\label{fig:ERvsEpsilon}
\end{figure}

The ergodic rate against the probabilistic interference constraint for different collision thresholds is illustrated in Fig. \ref{fig:ERvsEpsilon}. Increasing the maximum probability of violating the collision threshold improves $R$, the tradeoff, however, is the greater possibility of degrading the primary service operation which is deemed undesirable in practical scenarios. For more stringent collision limits, i.e., small $I^{th}$, increasing the maximum collision probability provides higher gains in ergodic rate. E.g., varying $\varepsilon$ from 0\% to 5\%, increases $R$ by 26.2\% and 15.4\% with $I^{th} = 5$ and $I^{th} = 7$, respectively; the gain diminishes quickly however as increasing $\varepsilon$ in high $I^{th}$ region implies a faster rate towards reaching a saturated performance limit.      

\begin{figure}
\includegraphics[width=.5\textwidth]{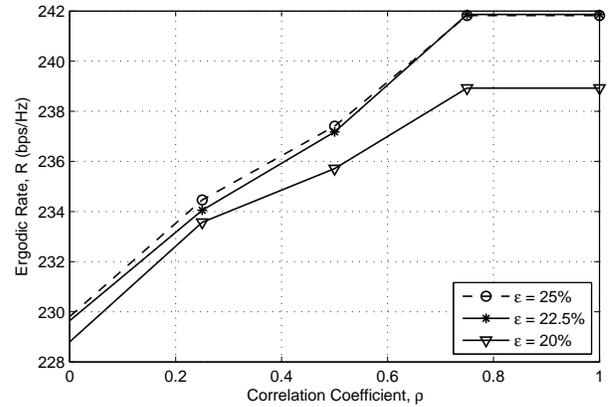} 
\caption{Ergodic rate against the correlation factor between estimation and error random variables for different probabilistic interference constraints. System parameters are: $P_t = 25$ Watts, $I^{th} = 20$ Watts.}
\label{fig:ERvsRho}
\end{figure}

System performance versus the correlation coefficient between the estimation and error random variables is depicted in Fig. \ref{fig:ERvsRho}. Higher $\rho$ correspond to greater accuracy of the estimation technique, with the ideal case of $\rho = 1$ where error variance is zero. Precise cross-link estimation is essential towards robust interference control, however, it typically implies more training symbols and thus increased signalling overhead.      


\section{Conclusions}

The ergodic transmission rate performance and limits of multi-user multi-band OFDMA CRs with underlay spectrum settings under imperfect cross-link knowledge was studied. We designed the downlink RA algorithm to maximize the cognitive network performance subject to satisfying average transmit and probabilistic peak interference power constraints. To compute the quantified loss in ergodic rate due to the added constraints, an approximated expression of the CRs' received SINR was developed. As a remedy to the imperfections associated with CSI and varying shared-spectrum environment, a stochastic interference management and exploitation technique was employed to confine the probability of collision to a pre-defined limit. The impact of parameters uncertainty on overall system performance was investigated through simulation results. Our proposed framework is an improvement in terms of practical feasibility and flexibility over the existing literature with perfect cross-link availability assumption, deterministic interference management policies, and single-user scenarios.

\bibliographystyle{IEEEtran}
\bibliography{IEEEabrv,myref}

\end{document}